\documentclass[]{spie}

\usepackage[]{graphicx}

\title{\Large\bf Braiding With Majorana Fermions}

\author{Louis H. Kauffman\supit{a} and Samuel J. Lomonaco Jr.2\supit{b}
\skiplinehalf
\supit{a} Department of Mathematics, Statistics and Computer Science  
(m/c 249), 851 South Morgan Street, University of Illinois at Chicago,
Chicago, Illinois 60607-7045, USA \\
\supit{b} Department of Computer Science and Electrical Engineering, University of
Maryland Baltimore County, 1000 Hilltop Circle, Baltimore, MD 21250, USA}
 
\authorinfo{Further author information: L.H.K.  E-mail: kauffman@uic.edu,
S.J.L. Jr.: E-mail: lomonaco@umbc.edu}
 
\begin{document} 
  \maketitle 

%%%%%%%%%%%%%%%%%%%%%%%%%%%%%%%%%%%%%%%%%%%%%%%%%%%%%%%%%%%%% 
\begin{abstract} 
This paper is an introduction to relationships between topology, quantum computing and the properties of fermions. In particular we study the remarkable unitary braid group representations associated with
Majorana fermions.  
 \end{abstract}

\keywords{braiding, fermions, majorana fermions, quantum process, quantum computing.}

\section{Introduction}
 In this paper we study a Clifford algebra generalization of the quaternions and its relationship with braid group representations related to Majorana fermions.  The Fibonacci model for topological quantum computing is  based on the fusion rules for a Majorana fermion. Majorana fermions can be seen not only in the structure of collectivies of electrons, as in the quantum Hall effect, but also in the structure of single electrons both by experiments with electrons in nanowires and also by the decomposition of the operator algebra for a fermion into a Clifford algebra generated by two Majorana operators. The purpose of this paper is to discuss these 
 braiding representations, important for relationships among physics, quantum information and topology. A new result in this paper is the Clifford Braiding Theorem of Section 3. This theorem shows that the 
 Majorana operators give rise to a particularly robust representation of the braid group that is then further represented to find the phases of the fermions under their exchanges in a plane space. The more robust 
 representation in our braiding theorem will be the subject of further work on our part.\\

\begin{figure}
     \begin{center}
     \begin{tabular}{c}
     \includegraphics[height=6cm]{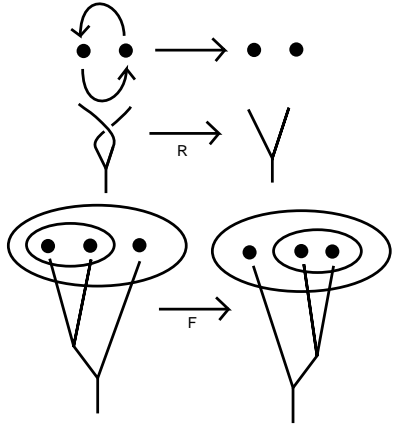}
     \end{tabular}
     \end{center}
     \caption{\bf Braiding Anyons }
     \label{Figure 1 }
     \end{figure}

\noindent{\bf Topological quantum computing.}
 For the sake of background, here is a very condensed presentation of how unitary representations of the braid group are constructed via topological quantum field theoretic methods, leading to the Fibonacci model and its generalizations.
One has a mathematical particle with label $P$
that can interact with itself to produce either itself labeled $P$ or itself
with the null label $*.$ We shall denote the interaction of two particles $P$ and $Q$ by the expression
$PQ,$ but it is understood that the ``value" of $PQ$ is the result of the interaction, and this may partake of a number of possibilities.
Thus for our particle $P$, we have that $PP$ may be equal to $P$ or to $*$ in a given situation.
When $*$ interacts with $P$ the result is always $P. $ When $*$ interacts with $*$ the result is always $*.$ One considers
process spaces where a row of particles labeled $P$ can successively
interact, subject to the restriction that the end result is $P.$ For example
the space $V[(ab)c]$ denotes the space of interactions of three particles
labeled $P.$ The particles are placed in the positions $a,b,c.$ Thus we
begin with $(PP)P.$ In a typical sequence of interactions, the first two $P$%
's interact to produce a $*,$ and the $*$ interacts with $P$ to produce $P.$ 
\[
(PP)P \longrightarrow (*)P \longrightarrow P. 
\]
\noindent In another possibility, the first two $P$'s interact to produce a $%
P,$ and the $P$ interacts with $P$ to produce $P.$ 
\[
(PP)P \longrightarrow (P)P \longrightarrow P. 
\]
It follows from this analysis that the space of linear combinations of
processes $V[(ab)c]$ is two dimensional. The two processes we have just
described can be taken to be the qubit basis for this space. One obtains
a representation of the three strand Artin braid group on $V[(ab)c]$ by
assigning appropriate phase changes to each of the generating processes. One
can think of these phases as corresponding to the interchange of the
particles labeled $a$ and $b$ in the association $(ab)c.$ The other operator
for this representation corresponds to the interchange of $b$ and $c.$ This
interchange is accomplished by a {\it unitary change of basis mapping} 
\[
F:V[(ab)c] \longrightarrow V[a(bc)]. 
\]
\noindent If 
\[
A:V[(ab)c] \longrightarrow V[(ba)c] 
\]
is the first braiding operator (corresponding to an interchange of the first
two particles in the association) then the second operator 
\[
B:V[(ab)c] \longrightarrow V[(ac)b] 
\]
is accomplished via the formula $B = F^{-1}RF$ where the $R$ in this formula
acts in the second vector space $V[a(bc)]$ to apply the phases for the
interchange of $b$ and $c.$ These issues are illustrated in Figure~\ref{Figure 1 }, where the parenthesization of the particles 
is indicated by circles and by also by trees. The trees can be taken to indicate patterns of particle interaction, where
two particles interact at the branch of a binary tree to produce the particle product at the root.  \bigbreak

     \bigbreak

In this scheme, vector spaces corresponding to associated strings of
particle interactions are interrelated by {\it recoupling transformations}
that generalize the mapping $F$ indicated above. A full representation of
the Artin braid group on each space is defined in terms of the local
interchange phase gates and the recoupling transformations. These gates and
transformations have to satisfy a number of identities in order to produce a
well-defined representation of the braid group. These identities were
discovered originally in relation to topological quantum field theory. In
our approach the structure of phase gates and recoupling
transformations arise naturally from the structure of the bracket model for
the Jones polynomial. Thus we obtain a knot-theoretic basis for topological
quantum computing. \bigbreak

In modeling the quantum Hall effect \cite{Wilczek,Fradkin,B1,B2}, the braiding of  quasi-particles (collective excitations) leads to non-trival
representations of the Artin braid group. Such particles are called {\it Anyons}. The braiding in these models is related to 
topological quantum field theory. \\

It is hoped that the mathematics we explain here will form a bridge between theoretical models of anyons and their applications to quantum computing.
We have summarized the recoupling approach in order to contrast it with the way that braiding of Majorana fermions occurs in the present paper via natural representations of Clifford algebras and also with the representations of the quaternions as $SU(2)$ to the Artin braid group. The recoupling theory is motivated by a hypothesis that one could observe the Majorana particles by watching their interactions and fusions.
It is possible that this is a correct hypothesis for the vortices of the quantum Hall effect. It is less likely to be the right framework for electrons in one-dimensional nano-wires. Nevertheless, these two modes of creatiing braid group representations intersect at the place where there are only three Majorana operators, generating a copy of the quaternions. It is possible that by handing Majorana fermions in triples in this way, one can work with  
the very rich braid group representations that are associated with the quaternions. We make the juxtaposition in this paper and intend further study.
\bigbreak

\noindent {\bf Acknowledgement.}  
Much of this paper is based upon our joint work in the papers \cite{TEQE,Spie,Teleport,QK1,QK2,QK3,QK4,QK5,BG,AnyonicTop,QCJP1,QCJP2,Fibonacci,KLogic}. We have woven  this work into the present paper in a form that is coupled with recent and previous work on relations with logic and with Majorana fermions.  
 
\bigbreak

\section{Braids}

\begin{figure}
     \begin{center}
     \begin{tabular}{c}
     \includegraphics[height=6cm]{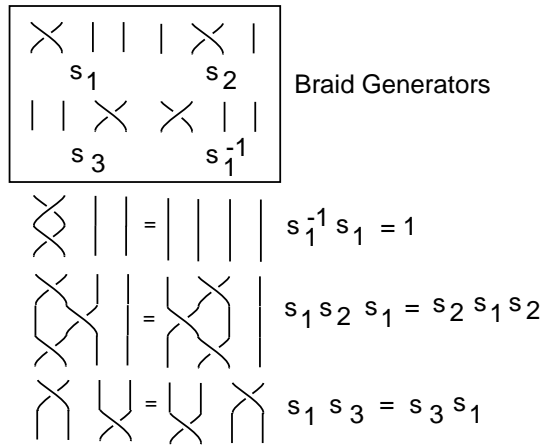}
     \end{tabular}
     \end{center}
     \caption{\bf Braid Generators  }
     \label{Figure 4 }
     \end{figure}

\begin{figure}
     \begin{center}
     \begin{tabular}{c}
     \includegraphics[height=6cm]{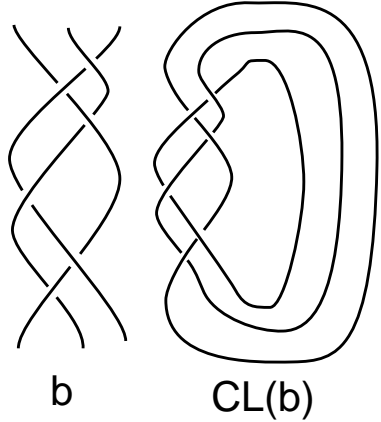}
     \end{tabular}
     \end{center}
     \caption{\bf Borromean Rings as a Braid Closure  }
     \label{Figure 6 }
     \end{figure}

A {\it braid} is an embedding of a collection of strands that have 
their ends in two rows of points that are set one above the other with respect to a choice of vertical. The strands are not
individually knotted and they are disjoint from one another. See Figure~\ref{Figure 4 }, and Figure~\ref{Figure 6 } for illustrations of braids and moves on braids. Braids can be 
multiplied by attaching the bottom row of one braid to the top row of the other braid. Taken up to ambient isotopy, fixing the endpoints, the braids form
a group under this notion of multiplication. In Figure~\ref{Figure 4 } we illustrate the form of the basic generators of the braid group, and the form of the
relations among these generators. Figure~\ref{Figure 6 } illustrates how to close a braid by attaching the top strands to the bottom strands by a collection of 
parallel arcs. A key theorem of Alexander states that every knot or link can be represented as a closed braid. Thus the theory of braids is critical to the 
theory of knots and links. Figure~\ref{Figure 6 } illustrates the famous Borromean Rings (a link of three unknotted loops such that any two of the loops are unlinked)
as the closure of a braid.
\bigbreak

     \bigbreak

     \bigbreak

Let $B_{n}$ denote the Artin braid group on $n$ strands.
We recall here that $B_{n}$ is generated by elementary braids $\{ s_{1}, \cdots ,s_{n-1} \}$
with relations 

\begin{enumerate}
\item $s_{i} s_{j} = s_{j} s_{i}$ for $|i-j| > 1$, 
\item $s_{i} s_{i+1} s_{i} = s_{i+1} s_{i} s_{i+1}$ for $i= 1, \cdots n-2.$
\end{enumerate}

\noindent See Figure~\ref{Figure 4 } for an illustration of the elementary braids and their relations. Note that the braid group has a diagrammatic
topological interpretation, where a braid is an intertwining of strands that lead from one set of $n$ points to another set of $n$ points.
The braid generators $s_i$ are represented by diagrams where the $i$-th and $(i + 1)$-th strands wind around one another by a single 
half-twist (the sense of this turn is shown in Figure~\ref{Figure 4 }) and all other strands drop straight to the bottom. Braids are diagrammed
vertically as in Figure~\ref{Figure 4 }, and the products are taken in order from top to bottom. The product of two braid diagrams is accomplished by
adjoining the top strands of one braid to the bottom strands of the other braid. 
\bigbreak 

In Figure~\ref{Figure 4 } we have restricted the illustration to the
four-stranded braid group $B_4.$ In that figure the three braid generators of $B_4$ are shown, and then the inverse of the
first generator is drawn. Following this, one sees the identities $s_{1} s_{1}^{-1} = 1$ 
(where the identity element in $B_{4}$ consists in  four vertical strands), 
$s_{1} s_{2} s_{1} = s_{2} s_{1}s_{2},$ and finally
$s_1 s_3 = s_3 s_1.$ 
\bigbreak

Braids are a key structure in mathematics. It is not just that they are a collection of groups with a vivid topological interpretation.
From the algebraic point of view the braid groups $B_{n}$ are important extensions of the symmetric groups $S_{n}.$ Recall that the 
symmetric group $S_{n}$ of all permutations of $n$ distinct objects has presentation as shown below.
\begin{enumerate}
\item $s_{i}^{2} = 1$ for  $i= 1, \cdots n-1,$
\item $s_{i} s_{j} = s_{j} s_{i}$ for $|i-j| > 1$, 
\item $s_{i} s_{i+1} s_{i} = s_{i+1} s_{i} s_{i+1}$ for $i= 1, \cdots n-2.$
\end{enumerate}
Thus $S_{n}$ is obtained from $B_{n}$ by setting the square of each braiding generator equal to one. We have an exact sequence of groups
$${1} \longrightarrow B_{n} \longrightarrow S_{n} \longrightarrow {1}$$ exhibiting the Artin braid group as an extension of the symmetric group.
\bigbreak

In the next sections we shall show how representations of the Artin braid group, rich enough to provide a dense set of transformations in the 
unitary groups, arise in relation to fermions and Majorana fermions. Braid groups are {\it in principle} fundamental to quantum computation and quantum information theory.
\bigbreak

\section {\bf Fermions, Majorana Fermions and Braiding}
\noindent{\bf Fermion Algebra.} Recall fermion algebra. One has fermion annihilation operators $\psi$ and their
conjugate creation operators $\psi^{\dagger}.$
One has $\psi^{2} = 0 = (\psi^{\dagger})^{2.}$
There is a fundamental commutation relation
$$\psi \psi^{\dagger} + \psi^{\dagger} \psi = 1.$$
If you have more than one of them say $\psi$ and $\phi$,
then they anti-commute:
$$\psi \phi = - \phi \psi.$$
The Majorana fermions $c$ satisfy $c^{\dagger} = c$ so that they
are their own anti-particles. There is a lot of interest in these as
quasi-particles and they are related to braiding and to topological
quantum computing.  A group of researchers  \cite{Kouwenhouven,Beenakker}
have found quasiparticle Majorana fermions in edge effects in nano-wires.
(A line of fermions could have a Majorana fermion happen non-locally from
one end of the line to the other.) The Fibonacci model that we discuss is also based on
Majorana particles, possibly related to collective electronic excitations. 
If $P$ is a Majorana fermion particle, then $P$ can interact with itself to either produce itself or to annihilate itself. This is the simple ``fusion algebra" for this particle. One can write
$P^2 = P + 1$ to denote the two possible self-interactions the particle $P.$
The patterns of interaction and braiding of such a particle $P$ give
rise to the Fibonacci model.\
\bigbreak

\noindent {\bf Majoranas make fermions.} Majoranas \cite{Majorana} are related to standard fermions as follows:
The algebra for Majoranas is $x = x^{\dagger}$ and $xy = -yx$ if $x$ and $y$ are
distinct Majorana fermions with  $x^{2}= 1$ and  $y^{2}= 1.$ Thus the operator algebra for a collection of Majorana particles is a Clifford algebra.
One can make a standard fermion from two Majoranas via
$$\psi = (x + iy)/2,$$
$$\psi^{\dagger} = (x -iy)/2.$$
Note, for example, that 
$$\psi^2 =  (x + iy)(x + iy)/4 = x^2 - y^2 + i(xy + yx) = 0+i0 = 0.$$
Similarly one can
mathematically make two Majoranas from any single fermion via
$$x = (\psi + \psi^{\dagger})/2$$
$$y = (\psi + \psi^{\dagger})/(2i).$$
This simple relationship between the fermion creation and annihilation algebra and an underlying Clifford algebra has long been a subject of speculation in physics. Only recently have experiments shown (indirect) 
evidence  \cite{Kouwenhouven} for Majorana fermions underlying the electron.\\

\noindent {\bf Braiding.} If you take a set of Majoranas
$$\{ c_1, c_2, c_3, \cdots , c_n \}$$
then there are natural braiding operators \cite{Ivanov,Kitaev} that act on the vector space with
these $c_k$ as the basis. The operators are mediated by algebra elements
$$\tau_{k} =(1 + c_{k+1} c_{k})/\sqrt{2},$$
$$\tau_{k}^{-1} = (1 - c_{k+1} c_{k})/\sqrt{2}2.$$
Then the braiding operators are
$$T_{k}: Span \{c_1,c_2,\cdots, ,c_n \} \longrightarrow Span \{c_1,c_2,\cdots, ,c_n \}$$
via 
$$T_{k}(x) = \tau_{k} x \tau_{k}^{-1}.$$
The braiding is simply:
$$T_{k}(c_{k}) = c_{k+1},$$
$$T_{k}(c_{k+1}) = - c_{k},$$
and $T_{k}$ is the identity otherwise.
This gives a very nice unitary representaton of the Artin braid group and
it deserves better understanding.
\bigbreak

\noindent That there is much more to this braiding is indicated by the following result.\\

\noindent {\bf Clifford Braiding Theorem.} Let $C$ be the Clifford algebra over the real numbers generated by linearly independent elements $\{ c_{1},c_{2}, \cdots c_{n}\}$ with 
$c_{k}^2 = 1$ for all $k$ and $c_{k}c_{l} = - c_{l}c_{k}$ for $k \ne l.$
Then the  algebra elements $\tau_{k} =(1 + c_{k+1} c_{k})/\sqrt{2},$ form a representation of the (circular) Artin braid group.
That is, we have $\{\tau_{1},\tau_{2}, \cdots \tau_{n-1}, \tau_{n} \}$ where $\tau_{k} =(1 + c_{k+1} c_{k})/\sqrt{2}$ for $1 \le k < n$ and $\tau_{n} =(1 + c_{1} c_{n})/\sqrt{2},$
and $\tau_{k}\tau_{k+1}\tau_{k} = \tau_{k+1}\tau_{k}\tau_{k+1}$ for all $k$ and  $\tau_{i}\tau_{j} = \tau_{j}\tau_{i}$ when $|i-j|>2.$  Note that each braiding generator
$\tau_{k}$ has order $8.$ Note also that we can formally write $\tau_{k} = exp(c_{k+1}c_{k} \pi/4).$\\

\noindent {\bf Proof.} Let $a_{k} = c_{k+1}c_{k}.$
Examine the following calculation:
$$\tau_{k}\tau_{k+1}\tau_{k} = (\frac{1}{2\sqrt{2}})(1 + a_{k+1})(1 + a_{k})(1 + a_{k+1})$$
$$ = (\frac{1}{2\sqrt{2}})(1 + a_{k}  + a_{k+1} + a_{k+1}a_{k})(1 + a_{k+1})$$
$$ = (\frac{1}{2\sqrt{2}})(1 + a_{k}  + a_{k+1} + a_{k+1}a_{k} + a_{k+1} + a_{k}a_{k+1}  + a_{k+1}a_{k+1} + a_{k+1}a_{k}a_{k+1})$$
$$ = (\frac{1}{2\sqrt{2}})(1 + a_{k}  + a_{k+1} + c_{k+2}c_{k} + a_{k+1} + c_{k}c_{k+2}  -1 -c_{k}c_{k+1})$$
$$ = (\frac{1}{2\sqrt{2}})(a_{k}  + a_{k+1} + a_{k+1} + c_{k+1}c_{k})$$
$$ = (\frac{1}{2\sqrt{2}})(2a_{k}  + 2a_{k+1})$$
$$ = (\frac{1}{\sqrt{2}})(a_{k}  + a_{k+1}).$$
Since the end result is symmetric under the interchange of $k$ and $k+1,$ we conclude that 
$$\tau_{k}\tau_{k+1}\tau_{k} = \tau_{k+1}\tau_{k}\tau_{k+1}.$$
Note that this braiding relation works circularly if we define $\tau_{n} =(1 + c_{1} c_{n})/\sqrt{2}.$
It is easy to see that $\tau_{i}\tau_{j} = \tau_{j}\tau_{i}$ when $|i-j|>2.$ This completes the proof. //\\

Undoubtedly, this representation of the (circular) Artin braid group is significant for the topological physics of Majorana fermions.
This part of the structure needs further study.\\

It is worth noting that a triple of Majorana fermions say $x,y,z$ gives rise
to a representation of the quaternion group. This is a generalization of
the well-known association of Pauli matrices and quaternions.
We have $x^2 = y^2 = z^2 = 1$ and, when different, they anti-commute.
Let $I = yx, J = zy, K = xz.$
Then $$I^2 = J^2 = K^2 = IJK = -1,$$ giving the quaternions.
The operators
$$A = (1/\sqrt{2})(1 + I)$$
$$B = (1/\sqrt{2})(1 + J)$$
$$C = (1/\sqrt{2})(1 + K)$$
braid one another: $$ABA = BAB,BCB =CBC, ACA = CAC.$$
This is a special case of the braid group representation described above
for an arbitrary list of Majorana fermions.
These braiding operators are entangling and so can be used for universal
quantum computation, but they give only partial topological quantum
computation due to the interaction with single qubit operators not
generated by them.
\bigbreak

Here is the derivation of the braiding relation in the quaternion case.
 $$ABA = (1/2 \sqrt{2})(1 + I)(1+ J)(1+I)$$
$$= (1/2 \sqrt{2})(1 + I + J + IJ)(1+I)$$
$$ =  (1/2 \sqrt{2})(1 + I + J + IJ + I + I^2 + JI + IJI)$$
$$=  (1/2 \sqrt{2})(1 + I + J + IJ + I  -1 - IJ + J)$$
$$ = (1/ \sqrt{2})(I + J).$$
The same form of computation yields 
$BAB= (1/ \sqrt{2})(J + I).$
And so $$ABA=BAB.$$
and so a natural braid group representation arises from the Majorana fermions.\\

\begin{figure}
     \begin{center}
     \begin{tabular}{c}
     \includegraphics[height=7cm]{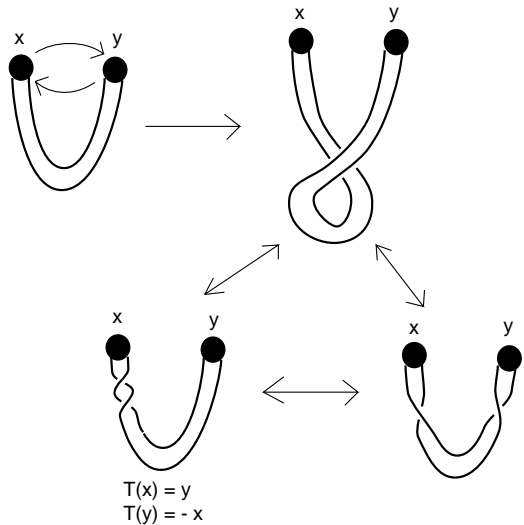}
     \end{tabular}
     \end{center}
     \caption{\bf Braiding Action on a Pair of Fermions}
     \label{ex}
     \end{figure}

These braiding operators can be seen to act on the vector space over the complex numbers that is spanned by the fermions $x, y, z.$ To see how this works, consider
$$s =  \frac{1 + yx}{\sqrt{2}},$$
$$T(p) = sps^{-1} = (\frac{1 + yx}{\sqrt{2}})p(\frac{1 - yx}{\sqrt{2}}),$$ 
and verify that 
$T(x) = y$ and $T(y) = -x.$ Now view Figure~\ref{ex} where we have illustrated a 
topological interpretation for the braiding of two fermions. In the topological interpretation the two fermions are connected by a flexible belt. On interchange, the belt becomes twisted  by $2 \pi.$ 
In the topological interpretation a twist of $2 \pi$ corresponds to a phase change of $-1.$ 
(For more information on this topological interpretation of $2 \pi$ rotation for fermions, see \cite{KP}.)
Without a further choice it is not evident which particle of the pair should receive the phase change. The topology alone tells us only the relative change of phase between the two particles. The Clifford algebra for Majorana fermions makes a specific choice in the matter and in this way fixes the representation of the braiding.
\bigbreak

     \bigbreak

A remarkable feature of this braiding representation of Majorana fermions is that it applies to give a representation of the $n$-strand braid group $B_{n}$ for any row of $n$ Majorana Fermions. It is not restricted to the 
quaternion algebra. Nevertheless, we shall now examine the braiding representations of the quaternions. These representations are very rich and can be used in situations (such as Fibonacci particles) involving particles that are their own anti-particles (analogous to the Majorana fermions underlying electrons). Such particles can occur in collectivities of electrons as in the quantum Hall effect. In such situations it is theorized that one can examine the local interaction properties of the Majorana particles and then the braidings associated to triples of them (the quaternion cases) can come into play very strongly. In the case of electrons in nano-wires, one at the present time must make do with long range correlations between ends of the wires and forgoe such local interactions. Nevertheless, it is the purpose of this paper to juxtapose the full story about three strand braid group represenations of the quaternions in the hope that this will lead to deeper understanding of the possibilities for even the electronic Majorana fermions.\\

\section{Braiding Operators and Universal Quantum Gates}
 
A key concept in the construction of quantum link invariants is
the association of a Yang-Baxter operator $R$ to each elementary crossing in a
link diagram. The operator $R$ is a linear mapping  
$$R\colon \ V\otimes V \longrightarrow V\otimes V$$ 
defined on the  $2$-fold tensor product of a vector space $V,$ generalizing the permutation of the factors
(i.e., generalizing a swap gate when $V$ represents one qubit). Such transformations are not 
necessarily unitary in 
topological applications. It is useful to understand
when they can be replaced by unitary transformations 
for the purpose of quantum 
computing. Such unitary $R$-matrices can be used to 
make unitary representations of the Artin braid group.
\bigbreak

\begin{figure}
     \begin{center}
     \begin{tabular}{c}
     \includegraphics[height=4cm]{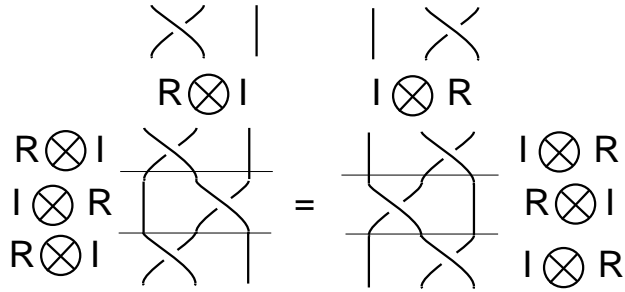}
     \end{tabular}
     \end{center}
     \caption{\bf The Yang-Baxter equation }
     \label{Figure 7 }
     \end{figure}

A solution to the Yang-Baxter equation, as described in the last 
paragraph is a matrix $R,$ regarded as a mapping of a
two-fold tensor product of a vector space
$V \otimes V$ to itself that satisfies the equation 

$$(R \otimes I)(I \otimes R)(R \otimes I) = 
(I \otimes R)(R \otimes I)(I \otimes R).$$ From the point of view of topology, the matrix $R$ 
is regarded as representing an elementary bit of braiding 
represented by one string
crossing over another. In Figure~\ref{Figure 7 } we have illustrated 
the braiding identity that corresponds to the Yang-Baxter equation.
Each braiding picture with its three input lines (below) 
and output lines (above) corresponds to a mapping of the three fold
tensor product of the vector space $V$ to itself, as required 
by the algebraic equation quoted above. The pattern of placement of the 
crossings in the diagram corresponds to the factors 
$R \otimes I$ and $I \otimes R.$ This crucial 
topological move has an algebraic
expression in terms of such a matrix $R.$  We need to study solutions of the Yang-Baxter equation that are unitary.
Then the $R$ matrix can be seen {\em either} as a braiding matrix 
{\em or} as a quantum gate in a quantum computer.

     \bigbreak

 \subsection{Universal Gates}
A {\em two-qubit gate} $G$ is a unitary  linear mapping $G:V \otimes V \longrightarrow V$ where $V$ is a two complex dimensional
vector space. We say that the gate $G$ is {\em universal for quantum computation} (or just {\em universal}) if $G$ together with 
local unitary transformations (unitary transformations from $V$ to $V$) generates all unitary transformations of the complex vector
space of dimension $2^{n}$ to itself. It is well-known \cite{N} that $CNOT$ is a universal gate. (On the standard basis,
$CNOT$ is the identity when the first qubit is $|0 \rangle $, and it flips the second qbit, leaving the first alone, when the first qubit is $|1 \rangle .$)
\bigbreak

\noindent A gate $G$, as above, is said to be {\em entangling} if there is a vector  
$$| \alpha \beta \rangle = | \alpha \rangle \otimes | \beta \rangle \in V \otimes V$$ such that 
$G | \alpha \beta \rangle$ is not decomposable as a tensor product of two qubits. Under these circumstances, one says that 
$G | \alpha \beta \rangle$ is {\em entangled}.
\bigbreak

\noindent In \cite{BB},  the Brylinskis
give a general criterion of $G$ to be universal. They prove that {\em a two-qubit gate $G$ is universal if and only if it is
entangling.} 
\bigbreak

\noindent {\bf Remark.} A two-qubit pure state $$|\phi \rangle = a|00 \rangle + b|01 \rangle + c|10 \rangle + d|11 \rangle$$
is entangled exactly when $(ad-bc) \ne 0.$ It is easy to use this fact to check when a specific matrix is, or is not, entangling.
\bigbreak

\noindent {\bf Remark.} There are many gates other than $CNOT$ that can be used as universal gates in the presence of local unitary 
transformations. Some of these are themselves topological (unitary solutions to the Yang-Baxter equation, see \cite{BG,BA}) and themselves generate
representations of the Artin braid group. Replacing $CNOT$ by a solution to the Yang-Baxter equation does not place the local unitary transformations as
part of the corresponding representation of the braid group. Thus such substitutions give only a partial solution to creating topological 
quantum computation. 
\bigbreak

\subsection{\bf Majorana Fermions Generate Universal Braiding Gates}
Recall that in Section 3 we showed how to construct braid group representations. Let  $T_{k}: V_{n} \longrightarrow V_{n}$
defined by $$T_{k}(v) = s_{k} v s_{k}^{-1}$$ be defined as in Section 3.  Note that $ s_{k}^{-1} = \frac{1}{\sqrt{2}}(1 - c_{k+1} c_{k}).$ It is then easy to verify that $$T_{k}(c_{k}) = c_{k+1},$$ $$T_{k}(c_{k+1}) = - c_{k}$$ and that $T_{k}$ is the identity otherwise.\\

For universality, take $n = 4$ and regard each $T_{k}$ as operating on $V \otimes V$ where $V$ is a single qubit space.  Then the braiding operators $T_{k}$ each satisfy the Yang-Baxter equation and are entangling operators, and so we have universal gates (in the presence of single qubit unitary operators) from Majorana fermions. If experimental work shows that Majorana fermions can be detected and controlled, then it is possible that quantum computers based on these topological unitary representations will be constructed.
\bigbreak

\section{$SU(2)$ Representations of the Artin Braid Group}
The purpose of this section is to determine all the representations of the three strand Artin braid group $B_{3}$ to the special unitary group $SU(2)$ and
concomitantly to the unitary group $U(2).$ One regards the groups $SU(2)$ and $U(2)$ as acting on a single qubit, and so $U(2)$ is usually regarded as the
group of local unitary transformations in a quantum information setting. If one is looking for a coherent way to represent all unitary transformations by
way of braids, then $U(2)$ is the place to start. Here we will show that there are many representations of the three-strand braid group
that generate a dense subset of $U(2).$ Thus it is a fact that local unitary transformations can be "generated by braids" in many ways.
\bigbreak

We begin with the structure of $SU(2).$ A matrix in $SU(2)$ has the form 
$$ M = 
\left( \begin{array}{cc}
z & w \\
-\bar{w} & \bar{z} \\
\end{array} \right),$$ where $z$ and $w$ are complex numbers, and $\bar{z}$ denotes the complex conjugate of $z.$ 
To be in $SU(2)$ it is required that $Det(M)=1$ and that $M^{\dagger} = M^{-1}$ where $Det$ denotes determinant, and $M^{\dagger}$ is the conjugate transpose of $M.$
Thus if
$z = a + bi$ and $w = c + di$ where $a,b,c,d$ are real numbers, and $i^2 = -1,$ then 
$$ M = 
\left( \begin{array}{cc}
a + bi & c + di \\
-c + di & a - bi \\
\end{array} \right)$$  with $a^2 + b^2 + c^2 + d^2 = 1.$ It is convenient to write
$$M =
a\left( \begin{array}{cc}
1 & 0 \\
0 & 1 \\
\end{array} \right) +
b\left( \begin{array}{cc}
i & 0\\
0 & -i \\
\end{array} \right) +
c\left( \begin{array}{cc}
0  & 1 \\
-1  & 0\\
\end{array} \right) +
d\left( \begin{array}{cc}
0 & i \\
i & 0 \\
\end{array} \right),$$ and to abbreviate this decomposition as
$$M = a + bI +cJ + dK$$
where 
$$ 1 \equiv
\left( \begin{array}{cc}
1 & 0 \\
0 & 1 \\
\end{array} \right),
I \equiv
\left( \begin{array}{cc}
i & 0\\
0 & -i \\
\end{array} \right),
J \equiv
\left( \begin{array}{cc}
0  & 1 \\
-1  & 0\\
\end{array} \right),
K \equiv
\left( \begin{array}{cc}
0 & i \\
i & 0 \\
\end{array} \right)$$ so that 
$$I^2 = J^2 = K^2 = IJK = -1$$ and 
$$IJ = K, JK=I, KI = J$$
$$JI = -K, KJ = -I, IK = -J.$$
The algebra of $1,I,J,K$ is called the {\it quaternions} after William Rowan Hamilton who discovered this algebra prior to the discovery of 
matrix algebra. Thus the unit quaternions are identified with $SU(2)$ in this way. We shall use this identification, and some facts about 
the quaternions to find the $SU(2)$ representations of braiding. First we recall some facts about the quaternions.

\begin{enumerate}
\item Note that if $q = a + bI +cJ + dK$ (as above), then $q^{\dagger} = a - bI - cJ - dK$ so that $qq^{\dagger} = a^2 + b^2 + c^2 + d^2 = 1.$
\item A general quaternion has the form $ q = a + bI + cJ + dK$ where the value of $qq^{\dagger} = a^2 + b^2 + c^2 + d^2,$ is not fixed to unity.
The {\it length} of $q$ is by definition $\sqrt{qq^{\dagger}}.$
\item A quaternion of the form $rI + sJ + tK$ for real numbers $r,s,t$ is said to be a {\it pure} quaternion. We identify the set of pure
quaternions with the vector space of triples $(r,s,t)$ of real numbers $R^{3}.$
\item Thus a general quaternion has the form $q = a + bu$ where $u$ is a pure quaternion of unit length and $a$ and $b$ are arbitrary real numbers.
A unit quaternion (element of $SU(2)$) has the addition property that $a^2 + b^2 = 1.$
\item If $u$ is a pure unit length quaternion, then $u^2 = -1.$ Note that the set of pure unit quaternions forms the two-dimensional sphere
$S^{2} = \{ (r,s,t) | r^2 + s^2 + t^2 = 1 \}$ in $R^{3}.$
\item If $u, v$ are pure quaternions, then $$uv = -u \cdot v + u \times v$$ whre $u \cdot v$ is the dot product of the vectors $u$ and $v,$ and 
$u \times v$ is the vector cross product of $u$ and $v.$ In fact, one can take the definition of quaternion multiplication as
$$(a + bu)(c + dv) = ac + bc(u) + ad(v) + bd(-u \cdot v + u \times v),$$ and all the above properties are consequences of this
definition. Note that quaternion multiplication is associative.
\item Let $g = a + bu$ be a unit length quaternion so that $u^2 = -1$ and $a = cos(\theta/2), b=sin(\theta/2)$ for a chosen angle $\theta.$
Define $\phi_{g}:R^{3} \longrightarrow R^{3}$ by the equation $\phi_{g}(P) = gPg^{\dagger},$ for $P$ any point in $R^{3},$ regarded as a pure quaternion.
Then $\phi_{g}$ is an orientation preserving rotation of $R^{3}$ (hence an element of the rotation group $SO(3)$). Specifically, $\phi_{g}$ is a rotation
about the  axis $u$ by the angle $\theta.$ The mapping $$\phi:SU(2) \longrightarrow SO(3)$$ is a two-to-one surjective map from the special unitary group to
the rotation group. In quaternionic form, this result was proved by Hamilton and by Rodrigues in the middle of the nineteeth century.
The specific formula for $\phi_{g}(P)$ as shown below:
$$\phi_{g}(P) = gPg^{-1} = (a^2 - b^2)P + 2ab (P \times u) + 2(P \cdot u)b^{2}u.$$
\end{enumerate}

We want a representation of the three-strand braid group in $SU(2).$ This means that we want a homomorphism $\rho: B_{3} \longrightarrow SU(2),$ and hence
we want elements $g = \rho(s_{1})$ and $h= \rho(s_{2})$ in $SU(2)$ representing the braid group generators $s_{1}$ and $s_{2}.$ Since $s_{1}s_{2}s_{1} =
s_{2}s_{1}s_{2}$ is the generating relation for $B_{3},$ the only requirement on $g$ and $h$ is that $ghg = hgh.$ We rewrite this relation as
$h^{-1}gh = ghg^{-1},$ and analyze its meaning in the unit quaternions.
\bigbreak

Suppose that $g = a + bu$ and $h=c + dv$ where $u$ and $v$ are unit pure quaternions so that $a^2 + b^2 = 1$ and $c^2 + d^2 = 1.$
then $ghg^{-1} = c +d\phi_{g}(v)$ and $h^{-1}gh = a + b\phi_{h^{-1}}(u).$ Thus it follows from the braiding relation that 
$a=c,$ $b= \pm d,$ and that $\phi_{g}(v) = \pm \phi_{h^{-1}}(u).$  However, in the case where there is a minus sign we have
$g = a + bu$ and $h = a - bv = a + b(-v).$ Thus we can now prove the following Theorem. 
\bigbreak

\noindent {\bf Theorem.} Let $u$ and $v$ be pure unit quaternions and $g = a + bu$ and $h=c + dv$ have unit length. Then (without loss of generality), the braid relation $ghg=hgh$ is true if
and only if
$h = a + bv,$ and $\phi_{g}(v) = \phi_{h^{-1}}(u).$ Furthermore, given that $g = a +bu$ and $h = a +bv,$ the condition $\phi_{g}(v) = \phi_{h^{-1}}(u)$
is satisfied if and only if $u \cdot v = \frac{a^2 - b^2}{2 b^2}$ when $u \ne v.$ If $u = v$ then $g = h$ and the braid relation is trivially
satisfied.
\bigbreak

\noindent {\bf Proof.} We have proved the first sentence of the Theorem in the discussion prior to its statement. Therefore assume that
$g = a +bu, h = a +bv,$ and $\phi_{g}(v) = \phi_{h^{-1}}(u).$ 
We have already stated the formula for $\phi_{g}(v)$ in the discussion about quaternions:
$$\phi_{g}(v) = gvg^{-1} = (a^2 - b^2)v + 2ab (v \times u) + 2(v \cdot u)b^{2}u.$$ By the same token, we have
$$\phi_{h^{-1}}(u) = h^{-1}uh = (a^2 - b^2)u + 2ab (u \times -v) + 2(u \cdot (-v))b^{2}(-v)$$
$$= (a^2 - b^2)u + 2ab (v \times u) + 2(v \cdot u)b^{2}(v).$$ Hence we require that
$$(a^2 - b^2)v + 2(v \cdot u)b^{2}u = (a^2 - b^2)u + 2(v \cdot u)b^{2}(v).$$ This equation is equivalent to
$$2(u \cdot v)b^{2} (u - v) = (a^2 - b^2)(u - v).$$
If $u \ne v,$ then this implies that $$u \cdot v = \frac{a^2 - b^2}{2 b^2}.$$
This completes the proof of the Theorem. //
\bigbreak

\noindent{\bf The Majorana Fermion Example.} Note the case of the theorem where
$$g = a +bu, h = a +bv.$$ Suppose that $u \cdot v = 0.$ Then the theorem tells us that we need 
$a^2 - b^2 = 0$ and since $a^2 +b^2 = 1,$ we conclude that $a = 1/\sqrt{2}$ and $b$ likewise.
For definiteness, then we have for the braiding generators (since $I$, $J$ and $K$ are mutually orthogonal) the three operators
$$A = \frac{1}{\sqrt{2}}(1 + I),$$
$$B =\frac{1}{\sqrt{2}}(1 + J),$$
$$C = \frac{1}{\sqrt{2}}(1 + K).$$
Each pair satisfies the braiding relation so that $ABA = BAB, BCB = CBC, ACA =CAC.$ We have already met this braiding triplet in our discussion of the construction of braiding operators from Majorana fermions in  Section 3. This shows (again) how close Hamilton's quaternions are to topology and how braiding is fundamental to the structure of fermionic physics.
\bigbreak

\noindent{\bf The Fibonacci  Example.} Let
$$g = e^{I\theta} = a + bI$$ where $a = cos(\theta)$ and $b = sin(\theta).$
Let $$h = a + b[(c^2 - s^2)I + 2csK]$$ where $c^2 + s^2 = 1$ and $c^2 - s^2 = \frac{a^2 - b^2}{2b^2}.$ Then we can rewrite $g$ and $h$ in matrix form
as the matrices $G$ and $H.$ Instead of writing the explicit form of $H,$ we write $H = FGF^{\dagger}$ where $F$ is an element of $SU(2)$ as shown below.

$$G =
\left( \begin{array}{cc}
e^{i\theta} & 0 \\
0 & e^{-i\theta} \\
\end{array} \right)$$

$$F =
\left( \begin{array}{cc}
ic & is \\
is & -ic \\
\end{array} \right)$$
This representation of braiding where one generator $G$ is a simple matrix of phases, while the other generator $H = FGF^{\dagger}$ is derived from $G$ by
conjugation by a unitary matrix, has the possibility for generalization to representations of braid groups (on greater than three strands) to $SU(n)$ or
$U(n)$ for 
$n$ greater than $2.$ In fact we shall see just such representations \cite{AnyonicTop} by using a version of topological quantum field theory.
The simplest example is given by 
$$g = e^{7 \pi I/10}$$
$$f = I\tau  + K \sqrt{\tau}$$
$$h = f g f^{-1}$$
where $\tau^{2} + \tau = 1.$
Then $g$ and $h$ satisfy $ghg=hgh$ and generate a representation of the three-strand braid group that is dense in $SU(2).$ We shall call this the 
{\it Fibonacci} representation of $B_{3}$ to $SU(2).$
\bigbreak

 At this point we can close this paper with the speculation that braid group representations such as this Fibonacci representation can be realized in the context of electrons in nano-wires.
 The formalism is the same as our basic Majorana representation. It has the form of a braiding operators of the form $$exp(\theta yx)$$ where $x$ and $y$ are Majorana operators and the angle 
 $\theta$ is not equal to $\pi/4$ as is required in the full Majorana representation. For a triple $\{x,y,z\}$ of Majorana operators, any quaternion representation is available. Note how this will effect the 
 conjugation representation: Let $T = r + s yx$ where $r$ and $s$ are real numbers with $r^2 + s^2 = 1$ (the cosine and sine of $\theta$), chosen so that a representation of the braid group is formed
 at the triplet (quaternion level). Then $T^{-1} = r - s yx$ and the reader can verify that 
 $$TxT^{-1} = (r^2 - s^2)x + 2rs y,$$
 $$TyT^{-1} =   (r^2 - s^2)y - 2rs x.$$
 Thus we see that the original fermion exchange occurs with $r=s$ and then the sign on $-2rs$ is the well-known sign change in the exchange of fermions. Here it is generalized to a more complex linear combination
 of the two particle/operators. It remains to be seen what is the meaning of this pattern at the level of Majorana particles in nano-wires.\\


\begin{thebibliography}{99}


\bibitem{BA}  R.J. Baxter.  Exactly Solved Models in Statistical Mechanics.  Acad. Press (1982).

\bibitem{Beenakker} C. W. J. Beenakker, Search for Majorana fermions in superconductors,
arXiv: 1112.1950.

\bibitem{B1}
N. E. Bonesteel, L. Hormozi, G. Zikos and S. H. Simon, Braid topologies for quantum computation,
 Phys. Rev. Lett. 95 (2005), no. 14, 140503, 4 pp. quant-ph/0505665.

\bibitem{B2}
S. H. Simon, N. E. Bonesteel, M. H. Freedman, N. Petrovic and L. Hormozi, Topological quantum computing with only one mobile quasiparticle,  Phys. Rev. Lett. 96 (2006), no. 7, 070503, 4 pp.,
quant-ph/0509175.


\bibitem{BB}
J. L. Brylinski and R. Brylinski, Universal quantum gates, in {\em Mathematics of Quantum Computation}, Chapman \& Hall/CRC Press, Boca Raton, Florida, 2002 (edited by R. Brylinski and G. Chen).


\bibitem{CKL} Chen, G., L. Kauffman, and S. Lomonaco, (eds.), "\textbf{%
Mathematics in Quantum Computation and Quantum Technology}," Chapman \&
Hall/CRC , (2007).



\bibitem{Fradkin}
E. Fradkin and P. Fendley, Realizing non-abelian statistics in time-reversal invariant systems, Theory Seminar, Physics Department, UIUC, 4/25/2005.


\bibitem{Ivanov} 
D. A. Ivanov, Non-abelian statistics of half-quantum vortices in $p$-wave superconductors, 
Phys. Rev. Lett. 86, 268 (2001). 


\bibitem{KL}
L.H. Kauffman, {\em Temperley-Lieb Recoupling Theory and Invariants of Three-Manifolds},
Princeton University Press, Annals Studies {\bf 114} (1994). 


\bibitem {KP}
L.H. Kauffman, {\em Knots and Physics}, World Scientific Publishers (1991), 
Second Edition (1993), Third Edition (2002), Fourth Edition (2012).

\bibitem{KLogic} L. H. Kauffman, Knot logic and topological quantum computing with Majorana fermions, in ``Logic and Algebraic Structures in Quantum Computing", edited by Chubb, Eskadarian and Harizanov,
Cambridge University Press (2016), pp. 223-335.

\bibitem{TEQE} 
L.H. Kauffman and S. J. Lomonaco Jr.,  Quantum entanglement and topological 
entanglement,  New Journal of Physics {\bf 4} (2002), 73.1--73.18 
(http://www.njp.org/).

\bibitem{Teleport}
L. H. Kauffman, Teleportation Topology, quant-ph/0407224, (in the Proceedings
of the  2004 Byelorus Conference on Quantum Optics),  {\it Opt. Spectrosc.} 9, 2005, 227-232.

 
\bibitem{Spie} 
L.H. Kauffman and S. J. Lomonaco Jr.,
Entanglement Criteria - Quantum and Topological, in 
{\em Quantum Information and Computation - Spie Proceedings, 21-22 April, 2003, 
Orlando, FL}, Donkor, Pinch and Brandt (eds.), Volume 5105, pp. 51--58.

\bibitem{QK1} L. H. Kauffman and S. J. Lomonaco Jr., Quantum knots, in {\it Quantum Information
and Computation II, Proceedings of Spie, 12 -14 April 2004} (2004), ed. by Donkor Pirich and Brandt, pp. 268-284.

\bibitem{QK2} S. J. Lomonaco and L.  H. Kauffman, Quantum Knots and Mosaics, 
Journal of Quantum Information Processing, Vol. 7, Nos. 2-3, (2008), pp. 85 - 115. 
http://arxiv.org/abs/0805.0339

 \bibitem{QK3} S. J. Lomonaco and L. H. Kauffman, Quantum knots and lattices, or a blueprint for quantum systems that do rope tricks. Quantum information science and its contributions to mathematics, 209Ð276, Proc. Sympos. Appl. Math., 68, Amer. Math. Soc., Providence, RI, 2010.
 
 \bibitem{QK4} S. J. Lomonaco and L. H. Kauffman, Quantizing braids and other mathematical structures: the general quantization procedure. In Brandt, Donkor, Pirich, editors, {\em Quantum Information and Comnputation IX - Spie Proceedings, April 2011}, Vol. 8057, of Proceedings of Spie, pp. 805702-1 to 805702-14, SPIE 2011.

\bibitem{QK5} L. H. Kauffman and S. J. Lomonaco, Quantizing knots groups and graphs. In Brandt, Donkor, Pirich, editors, {\em Quantum Information and Comnputation IX - Spie Proceedings, April
2011}, Vol. 8057, of Proceedings of Spie, pp. 80570T-1 to 80570T-15, SPIE 2011.

\bibitem{BG}
L. H. Kauffman and S. J. Lomonaco, Braiding Operators are Universal Quantum
Gates, New Journal of Physics 6 (2004) 134, pp. 1-39.

\bibitem{AnyonicTop}
Kauffman, Louis H.; Lomonaco, Samuel J., Jr. $q$-deformed spin networks, knot polynomials and anyonic topological quantum computation. J. Knot Theory Ramifications 16 (2007), no. 3, 267--332. 
 
\bibitem{SpinTop}
L. H. Kauffman and S. J. Lomonaco Jr.,
Spin Networks and Quantum Computation, in ``LIe Theory and Its Applications in Physics VII"
eds. H. D. Doebner and V. K. Dobrev, Heron Press, Sofia (2008), pp. 225 - 239.
 

\bibitem{QCJP1}
L. H. Kauffman, Quantum computing and the Jones polynomial, math.QA/0105255, in {\em Quantum Computation and Information}, S. Lomonaco, 
Jr. (ed.), AMS CONM/305, 2002, pp.~101--137.

\bibitem{QCJP2}
L. H. Kauffman and S. J. Lomonaco Jr.,
Quantum entanglement and topological entanglement. New J. Phys. 4 (2002), 73.1Ð73.18.

\bibitem{Fibonacci}
L. H. Kauffman and S. J. Lomonaco Jr., The Fibonacci Model and the Temperley-Lieb Algebra.
{\ it International J. Modern Phys. B}, Vol. 22, No. 29 (2008), 5065-5080.


\bibitem{Kouwenhouven} 
V. Mourik,K. Zuo, S. M. Frolov, S. R. Plissard, E.P.A.M. Bakkers, L.P. Kouwenhuven, Signatures of Majorana fermions in hybred superconductor-semiconductor devices,
arXiv: 1204.2792.

 
\bibitem{Kitaev}
A. Kitaev, Anyons in an exactly solved model and beyond,  Ann. Physics 321 (2006), no. 1, 2Ð111.
{\em arXiv.cond-mat/0506438 v1 17 June 2005}.


\bibitem{Majorana} E. Majorana, A symmetric theory of electrons and positrons, 
I Nuovo Cimento,{\bf 14} (1937), pp. 171-184.

\bibitem{N}
M. A. Nielsen and I. L. Chuang, ``	Quantum Computation and Quantum Information," Cambrige University Press, Cambridge (2000). 

\bibitem{Wilczek}
F. Wilczek, {\em Fractional Statistics and Anyon Superconductivity,} World Scientific Publishing Company (1990).



\end{thebibliography}
\end{document}